\documentclass[10pt,english,sort&compress, 1p]{elsarticle}
\usepackage[T1]{fontenc}
\usepackage[latin9]{inputenc}
\usepackage{geometry}
\usepackage{mathrsfs}
\usepackage{amsmath}
\usepackage{graphicx}
\usepackage{wasysym}

\makeatletter

\providecommand{\tabularnewline}{\\}
\newcommand{\lyxdot}{.}

\usepackage{lipsum}
\renewcommand\appendix{\par
  \setcounter{section}{0}%
  \setcounter{subsection}{0}%
  \setcounter{equation}{0}
  \gdef\thefigure{\@Alph\c@section.\arabic{figure}}%
  \gdef\thetable{\@Alph\c@section.\arabic{table}}%
  \gdef\thesection{\appendixname\ \@Alph\c@section}%
  \@addtoreset{equation}{section}%
  \gdef\theequation{\@Alph\c@section.\arabic{equation}}%
  \addtocontents{toc}{\string\let\string\numberline\string\tmptocnumberline}{}{}
}

\makeatother

\usepackage{babel}
\begin{document}
\begin{frontmatter}

\title{Time-step dependent force interpolation scheme for suppressing numerical
Cherenkov instability in relativistic particle-in-cell simulations}
\begin{abstract}
The WT scheme, a piecewise polynomial force interpolation scheme with
time-step dependency, is proposed in this paper for relativistic particle-in-cell
(PIC) simulations. The WT scheme removes the lowest order numerical
Cherenkov instability (NCI) growth rate for arbitrary time steps allowed
by the Courant condition. While NCI from higher order resonances is
still present, the numerical tests show that for smaller time steps,
the numerical instability grows much slower than using the optimal
time step found in previous studies. The WT scheme is efficient for
improving the quality and flexibility of relativistic PIC simulations.
\end{abstract}
\author[1,2]{Yingchao Lu} 
\ead{yclu@lanl.gov}
\address[1]{Theoretical Division, Los Alamos National Laboratory, Los Alamos, New Mexico, 87545, USA}
\address[2]{Department of Physics and Astronomy, Rice University, Houston, Texas 77005, USA}
\author[1]{Patrick Kilian}
\author[1]{Fan Guo}
\author[1]{Hui Li}
\author[2]{Edison Liang}

\end{frontmatter}

\section{Introduction}

The particle-in-cell (PIC) method\citep{PIC_Birdsall1991} is widely
used for the simulations of plasma dynamics ranging from Laser Plasma
Accelerators (LPAs) to collisionless astrophysical problems. In the
PIC method, quasi-particles are used to sample the phase-space distribution
of physical charged particles. The equations of motion of quasi-particles
are solved using a particle-push algorithm, e.g. Boris algorithm\citep{PIC_Birdsall1991}.
The electromagnetic field is defined on a grid, usually the staggered
Yee grid\citep{PIC_Yee1966}. The Lorentz force acting on a quasi-particle
is calculated by interpolating the electromagnetic field from nearby
grid points to the quasi-particle location using a force interpolation
scheme. The on-grid current density is calculated using a current
deposition scheme according to the quasi-particle motion and is used
to update the on-grid electromagnetic field. The PIC method can be
implemented without solving a Poisson equation for the electric potential
if one uses an exact charge conservation scheme. Although the exact
charge conservation current deposition scheme\citep{Esirkepov2001}
allows an arbitrary form-factor for quasi-particle, the most commonly
used form-factor is a B-spline function. Using B-splines has a few
advantages\citep{1988csup.book.....H}, including the easiness of
computation due to their polynomial nature, the smoothness of the
charge assigned to the grid as the particles move across the grid,
and the negligible fluctuations at long-range. If one requires momentum
to be conserved, then the force interpolation function should be identical
to the charge assignment function. Higher order B-spline functions
have better smoothness and long-range properties, but are more computationally
expensive.

Relativistic PIC simulations with drifting plasma beams are vulnerable
to an electromagnetic numerical instability known as the Numerical
Cherenkov instability (NCI)\citep{NCI_Godfrey1974}. This numerical
instability is caused by the resonance between two modes in the numerical
method: (1) the vacuum electromagnetic mode, which has a deviation
of the dispersion relation from the physical one, i.e. $\omega=ck$,
due to the discretization of Maxwell equations, (2) the drifting plasma
beam mode, which is dispersionless but has its aliasing beam modes\citep{Aliasing_Huang2016}.
This resonance is a numerical artifact and unphysical. It is desirable
to have an efficient numerical method which significantly suppresses
the NCI in order to improve the quality of relativistic PIC simulations.
An analytical expression for lowest order NCI growth rate was derived\citep{NCI_Xu2013}.
The numerically most unstable mode and its growth rate can be calculated
from the analytical expression without carrying out any numerical
experiments. It was found that in the momentum conserving scheme,
if one uses time step $\Delta t=\Delta x_{1}/(2c)$ for a drifting
plasma in $x_{1}$ direction where $\Delta x_{1}$ is the grid spacing
in $x_{1}$ direction, the lowest order NCI growth rate vanishes\citep{NCI_Vay2011}.

In this work, we propose a time-step dependent force interpolation
scheme which removes the lowest order NCI growth for a drifting plasma
in $x_{1}$ direction for arbitrary time step allowed by the Courant
condition, not just for $\Delta t=\Delta x_{1}/(2c)$. We call this
interpolation scheme the ``WT scheme'', which stands for ``\textbf{w}eighting
with \textbf{t}ime-step dependency'', or for the form of multidimensional
interpolation function having $W$'s and $T$'s as in Eq(\ref{eq:interpolation-EB-real-mod}).
The quasi-particle form-factor for charge assignment is still a B-spline
function and the electrostatic part of the self-force vanishes for
the WT scheme. The WT scheme recovers the momentum conserving scheme\citep{Interpolation_Sokolov2013,Interpolation_Brackbill2016}
for the case where the time step is $\Delta t=\Delta x_{1}/(2c)$
and the grid spacings in all directions are the same. In the WT scheme,
the lowest order NCI growth rate still vanishes if the spatial derivative
stencil in the Faraday equation is modified\citep{Blinne2018} while
the spatial derivative stencil in the Ampere equation is unmodified.
However, the asymptotic expression for NCI growth rate only shows
the lowest order growth rate. High order terms do contribute to the
actual numerical simulations, but are complicated to derive analytically.
On the other hand, the numerical instabilities evolve nonlinearly
after saturation in the actual simulations. Thus numerical experiments
are carried out to quantify the behavior of NCI in full PIC simulations.
The numerical tests show that the simulation using the WT scheme is
more numerically stable if the time step is reduced. The WT scheme
has little impact on the computational cost, and thus is potentially
more efficient compared to the spatial Fourier transform based methods,
such as the Pseudo-Spectral Time Domain (PSTD) algorithms\citep{PSTD_Godfrey2015}
for which the lowest order NCI resonance is removed by improving the
numerical phase velocity of the electromagnetic wave.

The remainder of this paper is organized as follows. In Sec \ref{sec:shape},
we derive the expression for the WT scheme. In Sec \ref{sec:Additional-considerations},
we describe a few additional considerations for relativistic PIC method.
The results for numerical experiments are given in Sec \ref{sec:Numerical-experiments}.

\section{WT scheme\label{sec:shape}}

For a three-dimensional electromagnetic PIC code with momentum conserving
(MC)\citep{Interpolation_Sokolov2013,Interpolation_Brackbill2016}
and exact charge conservation scheme\citep{Esirkepov2001} in Cartesian
coordinate, the electromagnetic field that is spatially interpolated
from grid point $\boldsymbol{x}_{g}=(x_{g,1},x_{g,2},x_{g,3})=(n_{1}\Delta x_{1},n_{2}\Delta x_{2},n_{3}\Delta x_{3})$
($n_{1}$, $n_{2}$, $n_{3}$ can be half-integer or integer depending
whether the component of electromagnetic field has a half-grid offset
in the $i$-th direction) to a particle position $\boldsymbol{x}=(x_{1},x_{2},x_{3})$
can be expressed as

\begin{equation}
E_{i}(\boldsymbol{x})=\begin{array}{ccc}
\sum_{n_{1},n_{2},n_{3}}W_{l}(\boldsymbol{x}-\boldsymbol{x}_{g})E_{i}(\boldsymbol{x}_{g}) & \qquad & B_{i}(\boldsymbol{x})=\sum_{n_{1},n_{2},n_{3}}W_{l}(\boldsymbol{x}-\boldsymbol{x}_{g})B_{i}(\boldsymbol{x}_{g})\end{array}\label{interpolation-EB-real-org}
\end{equation}
and the on-grid charge density of a quasi-particle is calculated from
the form-factor

\begin{equation}
\rho(\boldsymbol{x}_{g})=\frac{q}{V_{c}}W_{l}(\boldsymbol{x}-\boldsymbol{x}_{g})\label{eq:shape-real}
\end{equation}
where

\begin{equation}
W_{l}(\boldsymbol{x}-\boldsymbol{x}_{g})=W_{l}^{(1)}(x_{1}-x_{g,1})W_{l}^{(2)}(x_{2}-x_{g,2})W_{l}^{(3)}(x_{3}-x_{g,3})
\end{equation}
and $W_{l}^{(i)}$ is the $l$-th order B-spline with width $(l+1)\Delta x_{i}$
in $i$-th direction, and $V_{c}$ is the volume of a mesh cell, for
one-dimensional schemes, $V_{c}=\Delta x_{1}$, for two dimensions,
$V_{c}=\Delta x_{1}\Delta x_{2}$, and for three, $V_{c}=\Delta x_{1}\Delta x_{2}\Delta x_{3}$.
In the exact charge conservation scheme\citep{Esirkepov2001}, the
current density associated with the motion of a single quasi-particle
is the unique linear combination of the form-factor differences in
consistency with the discrete continuity equation. The Fourier transform
of the interpolation tensor in Eq(\ref{interpolation-EB-real-org})
is\citep{NCI_Xu2013}
\begin{equation}
\begin{array}{ccc}
S_{E1}=s_{l,1}s_{l,2}s_{l,3}\eta_{1} &  & S_{B1}=\cos(\omega^{\prime}\Delta t/2)s_{l,1}s_{l,2}s_{l,3}\eta_{2}\eta_{3}\\
S_{E2}=s_{l,1}s_{l,2}s_{l,3}\eta_{2} & \qquad & S_{B2}=\cos(\omega^{\prime}\Delta t/2)s_{l,1}s_{l,2}s_{l,3}\eta_{1}\eta_{3}\\
S_{E3}=s_{l,1}s_{l,2}s_{l,3}\eta_{3} &  & S_{B3}=\cos(\omega^{\prime}\Delta t/2)s_{l,1}s_{l,2}\tau_{l,3}\eta_{1}\eta_{2}
\end{array}\label{eq:shapeEB-org}
\end{equation}
where the factor $\eta_{i}=(-1)^{\nu_{i}}$ is multiplied when the
electromagnetic field has a half-grid offset in the $i$-th direction,
and
\begin{equation}
s_{l,i}=\bigg(\frac{\sin(k_{i}^{\prime}\Delta x_{i}/2)}{k_{i}^{\prime}\Delta x_{i}/2}\bigg)^{l+1}\label{eq:s_l}
\end{equation}
and the aliasing frequency and wave vectors with aliasing orders $(\mu,\nu_{1},\nu_{2},\nu_{3})$
are 
\begin{equation}
\begin{array}{ccc}
\omega^{\prime}=\omega+\mu\frac{2\pi}{\Delta t},\quad\mu=0,\pm1,\pm2,\dots & \qquad\qquad & k_{i}^{\prime}\end{array}=k_{i}+\nu_{i}\frac{2\pi}{\Delta x_{i}},\quad\nu_{i}=0,\pm1,\pm2,\dots\label{eq:omega_k_aliasing}
\end{equation}
There is one momentum conserving interpolation tensor for each $l$,
and we call it MC$l$.

Derived in Ref. \citep{NCI_Xu2013}, the asymptotic expression for
NCI growth rate for a cold drifting plasma beam traveling in $x_{1}$
direction with an ultra-relativistic speed $v_{1}\to c$ is

\begin{equation}
\Gamma=\frac{\sqrt{3}}{2}\bigg|\frac{\omega_{p}^{2}c^{2}S_{J1}\{(S_{B3}\xi_{0}-S_{E2}[k]_{B1}c)[k]_{E2}k_{2}+(S_{B2}\xi_{0}-S_{E3}[k]_{B1}c)[k]_{E3}k_{3}\}}{2\xi_{0}^{2}\xi_{1}}\bigg|^{1/3}\label{eq:NCI-Xu2013}
\end{equation}
where $S_{Ji}$ is the interpolation tensor for the current density
after Fourier transformation\citep{NCI_Xu2013}, and $\omega_{p}=\sqrt{4\pi q^{2}n_{e}/(\gamma_{0}m_{e})}$
is the relativistic plasma frequency, and the bulk Lorentz factor
is $\gamma_{0}=1/\sqrt{1-\beta^{2}}=1/\sqrt{1-v_{1}^{2}/c^{2}}$,
and the finite difference operators are
\begin{equation}
[\omega]=\frac{\sin(\omega\Delta t/2)}{\Delta t/2},\qquad[k]_{Ei}=A_{i}\frac{\sin(k_{i}\Delta x_{i}/2)}{\Delta x_{i}/2},\qquad[k]_{Bi}=\frac{\sin(k_{i}\Delta x_{i}/2)}{\Delta x_{i}/2}
\end{equation}
where $[k]_{Ei}$ or $A_{i}$ depends on the spatial derivative stencil
in Faraday's equation, and $[k]_{Bi}$ is related to the spatial derivative
stencil in Ampere's equation which is unmodified from the standard
Yee scheme, and

\begin{equation}
\xi_{0}=\frac{\sin(k_{1}^{\prime}c\Delta t/2)}{\Delta t/2},\qquad\xi_{1}=\cos(k_{1}^{\prime}c\Delta t/2)
\end{equation}
For MC scheme where $S_{E}$ and $S_{B}$ are given by Eq(\ref{eq:shapeEB-org}),
the lowest order NCI growth rate given by Eq(\ref{eq:NCI-Xu2013})
depends on the time step and only vanishes for $\Delta t=\Delta x_{1}/(2c)$.
In order to remove the time-step dependency of the NCI growth rate
given by Eq(\ref{eq:NCI-Xu2013}), we propose the WT scheme, where
we modify Eq(\ref{eq:shapeEB-org}) to the following form
\begin{equation}
\begin{array}{ccc}
S_{E1}=s_{l,1}\tau_{l,2}\tau_{l,3}\eta_{1} &  & S_{B1}=\cos(\omega^{\prime}\Delta t/2)\tau_{l,1}s_{l,2}s_{l,3}\eta_{2}\eta_{3}\\
S_{E2}=\tau_{l,1}s_{l,2}\tau_{l,3}\eta_{2} & \qquad & S_{B2}=\cos(\omega^{\prime}\Delta t/2)s_{l,1}\tau_{l,2}s_{l,3}\eta_{1}\eta_{3}\\
S_{E3}=\tau_{l,1}\tau_{l,2}s_{l,3}\eta_{3} &  & S_{B3}=\cos(\omega^{\prime}\Delta t/2)s_{l,1}s_{l,2}\tau_{l,3}\eta_{1}\eta_{2}
\end{array}\label{eq:shapeEB}
\end{equation}
where
\begin{equation}
\tau_{l,i}=\bigg(\frac{\sin(k_{i}^{\prime}\Delta x_{i}/2)}{k_{i}^{\prime}\Delta x_{i}/2}\bigg)^{l}\bigg(\frac{\sin(k_{i}^{\prime}c\Delta t)}{k_{i}^{\prime}c\Delta t}\bigg)\label{eq:tau_l}
\end{equation}
For the interpolation tensor in Eq(\ref{eq:shapeEB}), the expression
for lowest order NCI growth rate vanishes for arbitrary time step
$\Delta t$ as shown in \ref{sec:NCI-growth-rate}. For $\Delta t=\Delta x_{i}/(2c)$,
we have $\tau_{l,i}=s_{l,i}$, which recovers the MC scheme.

To get the interpolation function in real space, we calculate the
inverse Fourier transform of $s_{l,i}$ and $\tau_{l,i}$. We constrain
our discussion to $\Delta t\le\Delta x_{i}/(2c)$, because for $\Delta t>\Delta x_{i}/(2c)$
the width of the interpolation function becomes large and more grid
points are needed for interpolation. The inverse Fourier transform
of $s_{l,i}$ is simply the $(l+1)$-th order B-spline function $W_{l}^{(i)}$.
The interpolation function corresponding to $\tau_{1,i}$ is
\begin{equation}
T_{1}^{(i)}(\tilde{x}_{i})=\mathcal{F}^{-1}(\tau_{1,i})=\begin{cases}
\frac{1+2\Delta\tilde{t}_{i}-2|\tilde{x}_{i}|}{4\Delta\tilde{t}_{i}} & \mathrm{if}\ \frac{1}{2}-\Delta\tilde{t}_{i}<|\tilde{x}_{i}|\le\frac{1}{2}+\Delta\tilde{t}_{i}\\
1 & \mathrm{if}\ |\tilde{x}_{i}|\le\frac{1}{2}-\Delta\tilde{t}_{i}\\
0 & \mathrm{otherwise}
\end{cases}
\end{equation}
where $\mathcal{F}^{-1}$ is the inverse Fourier transformation, $\Delta\tilde{t}_{i}=c\Delta t/\Delta x_{i}$,
and $\tilde{x}=(x_{i}-x_{g,i})/\Delta x_{i}$ is the normalized coordinate
difference between the particle and the grid point. The interpolation
function corresponding to $\tau_{2,i}$ is 
\begin{equation}
T_{2}^{(i)}(\tilde{x}_{i})=\mathcal{F}^{-1}(\tau_{2,i})=\begin{cases}
\frac{(\Delta\tilde{t}_{i}+1-|\tilde{x}_{i}|)^{2}}{4\Delta\tilde{t}_{i}} & \mathrm{if}\ 1-\Delta\tilde{t}_{i}<|\tilde{x}_{i}|\le1+\Delta\tilde{t}_{i}\\
1-|\tilde{x}_{i}| & \mathrm{if}\ \Delta\tilde{t}_{i}<|\tilde{x}_{i}|\le1-\Delta\tilde{t}_{i}\\
\frac{2\Delta\tilde{t}_{i}-\Delta\tilde{t}_{i}^{2}-\tilde{x}_{i}^{2}}{2\Delta\tilde{t}_{i}} & \mathrm{if}\ |\tilde{x}_{i}|\le\Delta\tilde{t}_{i}\\
0 & \mathrm{otherwise}
\end{cases}
\end{equation}
The interpolation function corresponding to $\tau_{3,i}$ is
\begin{equation}
T_{3}^{(i)}(\tilde{x}_{i})=\mathcal{F}^{-1}(\tau_{3,i})=\begin{cases}
\frac{(3+2\Delta\tilde{t}_{i}-2|\tilde{x}_{i}|)^{3}}{96\Delta\tilde{t}_{i}} & \mathrm{if}\ \frac{3}{2}-\Delta\tilde{t}_{i}<|\tilde{x}_{i}|\le\frac{3}{2}+\Delta\tilde{t}_{i}\\
\frac{4\Delta\tilde{t}_{i}^{2}+3(3-2|\tilde{x}_{i}|)^{2}}{24} & \mathrm{if}\ \frac{1}{2}+\Delta\tilde{t}_{i}<|\tilde{x}_{i}|\le\frac{3}{2}-\Delta\tilde{t}_{i}\\
\frac{-8\Delta\tilde{t}_{i}^{3}-36\Delta\tilde{t}_{i}^{2}(1-2|\tilde{x}_{i}|)-3(1-2|\tilde{x}_{i}|)^{3}+6\Delta\tilde{t}_{i}(15-12|\tilde{x}_{i}|-4\tilde{x}_{i}^{2})}{96\Delta\tilde{t}_{i}} & \mathrm{if}\ \frac{1}{2}-\Delta\tilde{t}_{i}<|\tilde{x}_{i}|\le\frac{1}{2}+\Delta\tilde{t}_{i}\\
\frac{9-4\Delta\tilde{t}_{i}^{2}-12\tilde{x}_{i}^{2}}{12} & \mathrm{if}\ |\tilde{x}_{i}|\le\frac{1}{2}-\Delta\tilde{t}_{i}\\
0 & \mathrm{otherwise}
\end{cases}
\end{equation}
The interpolation function corresponding to $\tau_{4,i}$ is
\begin{equation}
T_{4}^{(i)}(\tilde{x}_{i})=\mathcal{F}^{-1}(\tau_{4,i})=\begin{cases}
\frac{(\Delta\tilde{t}_{i}+2-|\tilde{x}_{i}|)^{4}}{48\Delta t_{i}} & \mathrm{if}\ 2-\Delta\tilde{t}_{i}<|\tilde{x}_{i}|\le2+\Delta\tilde{t}_{i}\\
\frac{(2-|\tilde{x}_{i}|)[(2-|\tilde{x}_{i}|)^{2}+\Delta\tilde{t}_{i}^{2}]}{6} & \mathrm{if}\ 1+\Delta\tilde{t}_{i}<|\tilde{x}_{i}|\le2-\Delta\tilde{t}_{i}\\
\frac{-(1-|\tilde{x}_{i}|)^{4}+2\Delta\tilde{t}_{i}(6-6|\tilde{x}_{i}|+|\tilde{x}_{i}|^{3})-6\Delta\tilde{t}_{i}^{2}(1-|\tilde{x}_{i}|)^{2}+2\Delta\tilde{t}_{i}^{3}|\tilde{x}_{i}|-\Delta\tilde{t}_{i}^{4}}{12\Delta\tilde{t}_{i}} & \mathrm{if}\ 1-\Delta\tilde{t}_{i}<|\tilde{x}_{i}|\le1+\Delta\tilde{t}_{i}\\
\frac{4-6\tilde{x}_{i}^{2}+3|\tilde{x}_{i}|^{3}-\Delta\tilde{t}_{i}^{2}(2-3|\tilde{x}_{i}|)}{6} & \mathrm{if}\ \Delta\tilde{t}_{i}<|\tilde{x}_{i}|\le1-\Delta\tilde{t}_{i}\\
\frac{3\tilde{x}_{i}^{4}+\Delta\tilde{t}_{i}(16-24\tilde{x}_{i}^{2})+18\Delta\tilde{t}_{i}^{2}\tilde{x}_{i}^{2}-8\Delta\tilde{t}_{i}^{3}+3\Delta\tilde{t}_{i}^{4}}{24\Delta\tilde{t}_{i}} & \mathrm{if}\ |\tilde{x}_{i}|\le\Delta\tilde{t}_{i}\\
0 & \mathrm{otherwise}
\end{cases}
\end{equation}
The width of $T_{l}^{(i)}$ is $l\Delta x_{i}+2c\Delta t$, which
decreases as the time step decreases. There is one WT scheme interpolation
tensor for each $l$, and we call it WT$l$. The full interpolation
form for electromagnetic field in WT$l$ scheme is

\begin{equation}
\begin{aligned}E_{1}(\boldsymbol{x}) & =\sum_{n_{1},n_{2},n_{3}}W_{l}^{(1)}(x_{1}-x_{g,1})T_{l}^{(2)}(x_{2}-x_{g,2})T_{l}^{(3)}(x_{3}-x_{g,3})E_{1}(\boldsymbol{x}_{g})\\
E_{2}(\boldsymbol{x}) & =\sum_{n_{1},n_{2},n_{3}}T_{l}^{(1)}(x_{1}-x_{g,1})W_{l}^{(2)}(x_{2}-x_{g,2})T_{l}^{(3)}(x_{3}-x_{g,3})E_{2}(\boldsymbol{x}_{g})\\
E_{3}(\boldsymbol{x}) & =\sum_{n_{1},n_{2},n_{3}}T_{l}^{(1)}(x_{1}-x_{g,1})T_{l}^{(2)}(x_{2}-x_{g,2})W_{l}^{(3)}(x_{3}-x_{g,3})E_{3}(\boldsymbol{x}_{g})\\
B_{1}(\boldsymbol{x}) & =\sum_{n_{1},n_{2},n_{3}}T_{l}^{(1)}(x_{1}-x_{g,1})W_{l}^{(2)}(x_{2}-x_{g,2})W_{l}^{(3)}(x_{3}-x_{g,3})B_{1}(\boldsymbol{x}_{g})\\
B_{2}(\boldsymbol{x}) & =\sum_{n_{1},n_{2},n_{3}}W_{l}^{(1)}(x_{1}-x_{g,1})T_{l}^{(2)}(x_{2}-x_{g,2})W_{l}^{(3)}(x_{3}-x_{g,3})B_{2}(\boldsymbol{x}_{g})\\
B_{3}(\boldsymbol{x}) & =\sum_{n_{1},n_{2},n_{3}}W_{l}^{(1)}(x_{1}-x_{g,1})W_{l}^{(2)}(x_{2}-x_{g,2})T_{l}^{(3)}(x_{3}-x_{g,3})B_{3}(\boldsymbol{x}_{g})
\end{aligned}
\label{eq:interpolation-EB-real-mod}
\end{equation}
And the on-grid charge density of a quasi-particle in the WT scheme
is still given by Eq(\ref{eq:shape-real}), which can be inserted
in the derivation of the current deposition in an exact charge conserving
scheme\citep{Esirkepov2001}. The combination of Eq(\ref{eq:shape-real})
with Eq(\ref{eq:interpolation-EB-real-mod}) has zero self-force under
certain condition as shown in \ref{sec:self-force}. The WT scheme
allows flexibility in the choice of the time step, because the asymptotic
expression for NCI growth rate vanishes for arbitrary time step $\Delta t$,
not just for $\Delta t=\Delta x_{1}/(2c)$ as found in previous studies\citep{NCI_Xu2013,NCI_Vay2011}.

\section{Additional considerations\label{sec:Additional-considerations}}

A few additional considerations in relativistic PIC method are discussed
in this section.

\textit{Maxwell solvers}: A fully explicit Maxwell solver is usually
more computationally efficient than FFT-based or implicit solvers.
The fully explicit Maxwell solvers in Ref. \citep{Blinne2018}, which
modify the spatial derivative stencil in Faraday's equation and keep
the spatial derivative stencil in Ampere's equation, are compatible
with the charge conserving deposition scheme\citep{Esirkepov2001}.
By choosing the coefficients for the stencil in Faraday's equation,
the dispersion error can be fourth order, i.e. $\omega/(ck)=1+\mathcal{O}(k\Delta x)^{4}$
as $k\Delta x\to0$, as shown in  \ref{sec:dispersion-error}, while
generally the dispersion error for most Maxwell solvers is second
order, i.e. $\omega/(ck)=1+\mathcal{O}(k\Delta x)^{2}$.

\textit{Relativistic pseudo-particle loading}: For loading pseudo-particles
with relativistic drifting distribution, a sampling of the distribution
function in the co-moving frame is usually performed and transformed
into the simulation frame\citep{Zenitani2015}. Taking the volume
transform between two frames into account is significant. Failing
to do so can cause error in particle loading for relativistic distributions.
We write down the method for loading particles with arbitrary boost
velocities in \ref{subsec:load-distribution}.

\textit{Ultra-relativistic scaling}: The scaling relations can be
used for Lorentz factor scaling of the ultra-relativistic PIC simulations.
The simulation results obtained for one value of $\gamma_{0}$ can
be scaled to get the results for other values of $\gamma_{0}$, as
long as $\gamma_{0}$ is large and the initial and boundary conditions
of the dimensionless equations do not depend on $\gamma_{0}$, where
$\gamma_{0}$ is the characteristic Lorentz factor of the system.

\textit{Partially skipping calculation}: Current deposition and the
particle momentum update can be skipped in the unperturbed plasma
flow region where it is known to follow pure drift motion and be absent
of physical instabilities. If the initial perturbed region is $\mathscr{\mathscr{A}}(t=0)=\{(x,y,z)|(x,y,z)\in\mathscr{\mathscr{A}}_{0}\}$,
then the perturbed region for a later time $t>0$ is $\mathscr{\mathscr{A}}(t)=\{(x_{t},y_{t},z_{t})|(x_{t}-x_{0})^{2}+(y_{t}-y_{0})^{2}+(y_{t}-y_{0})^{2}<c^{2}t^{2},\ (x_{0},y_{0},z_{0})\in\mathscr{\mathscr{A}}_{0}\}$.
This kind of skipping not only prevents numerical instabilities from
growing, but also reduces the computational cost with the aid of dynamical
load balancing. For example, in the PIC simulations for relativistic
shock\citep{SHOCK_Sironi2013}, the current deposition and the particle
momentum update can be skipped in the upstream flow. Alternatively,
one can use the expanding box in a setup with simple geometry\citep{SHOCK_Sironi2013}.

\section{Numerical experiments\label{sec:Numerical-experiments}}

\begin{table}
\caption{Parameters for the test problem of drifting pair plasma.\label{tab:Parameters-for-test1}}

\hfill{}%
\begin{tabular}{|c|c|}
\hline 
domain size & $L_{x}=16d_{e}$, $L_{y}=8d_{e}$\tabularnewline
\hline 
boundary condition & periodic in both $x$ and $y$\tabularnewline
\hline 
number of cells & $N_{x}=256$, $N_{y}=128$\tabularnewline
\hline 
pseudo-particles per cell & $N_{\mathrm{PPC}}=64$ (32 for each species)\tabularnewline
\hline 
drift Lorentz factor & $\gamma_{0}=1000$\tabularnewline
\hline 
temperature & $k_{B}T_{e}=k_{B}T_{i}=0.01m_{e}c^{2}$\tabularnewline
\hline 
time step & $\Delta t/\Delta t_{\mathrm{CFL}}=\Delta t/\big[\Delta x/(\sqrt{2}c)\big]=0.1,0.3,0.5,0.7,1/\sqrt{2}$\tabularnewline
\hline 
\end{tabular}\hfill{}
\end{table}

\begin{figure}
\hfill{}\includegraphics[scale=0.42]{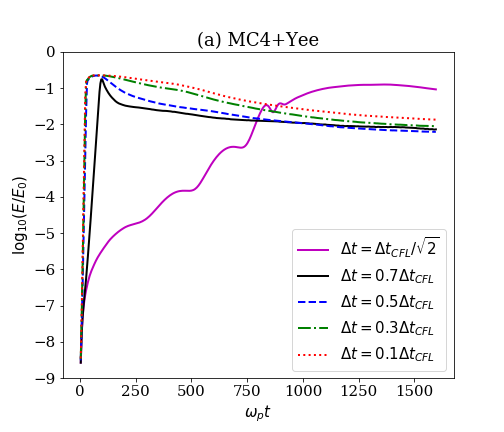}\includegraphics[scale=0.42]{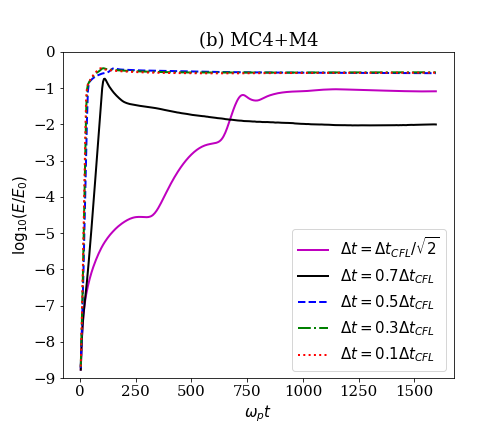}\hfill{}

\hfill{}\includegraphics[scale=0.42]{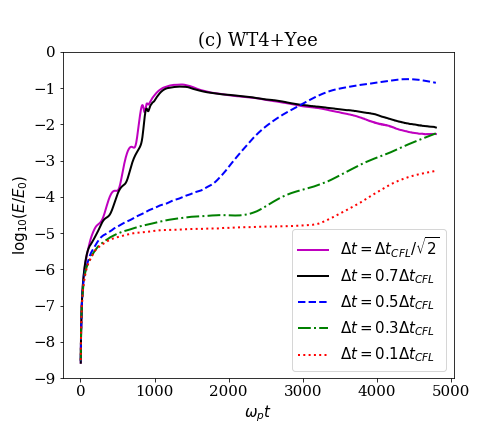}\includegraphics[scale=0.42]{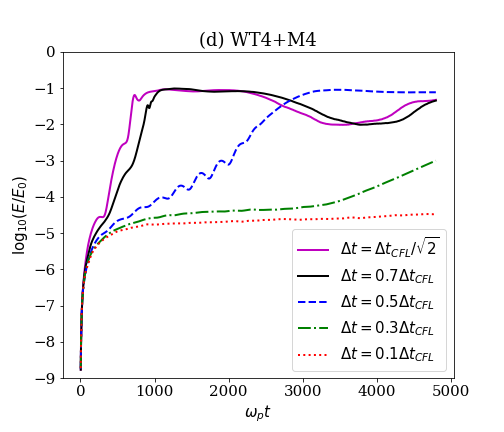}\hfill{}

\hfill{}\includegraphics[scale=0.42]{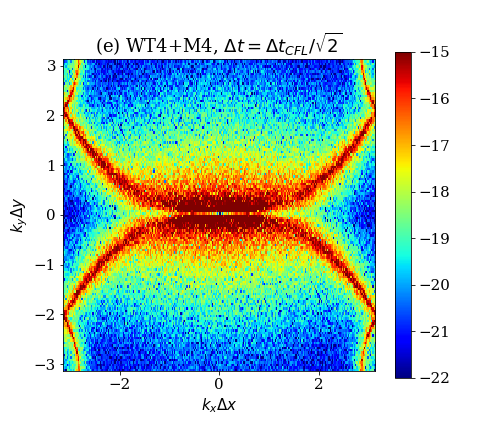}\includegraphics[scale=0.42]{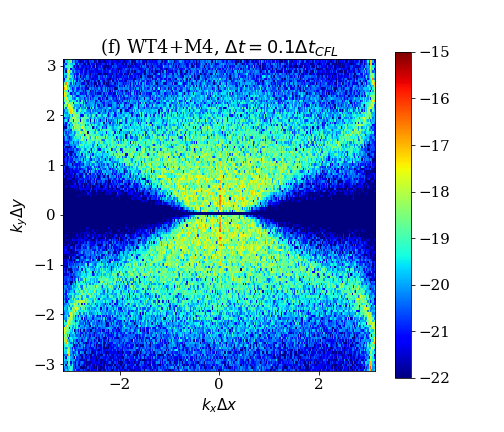}\hfill{}

\caption{Results for the test problem of drifting pair plasma. (a) The growth
history of the fraction of the total electromagnetic energy $E/E_{0}$
for MC4 interpolation scheme and Yee solver. (b) Same as (a) but for
MC4 interpolation scheme and M4 solver. (c) Same as (a) but for WT4
interpolation scheme and Yee solver. Note that the time axis is different
from (a). (d) Same as (a) but for WT4 interpolation scheme and M4
solver. Note that the time axis is different from (a) and (b). (e)
$\log(|\mathrm{FFT}(B_{z})|$) for WT4 scheme and M4 solver, and $\Delta t=\Delta t_{\mathrm{CFL}}/\sqrt{2}$.
The color bar is the logarithm of field energy in arbitrary units.
The simulation time for this frame is $t=1080\omega_{pe}^{-1}$. (f)
Same as (a) but with $\Delta t=0.1\Delta t_{\mathrm{CFL}}$.\label{fig:Results-for-test1}}
\end{figure}

We carry out two-dimensional numerical experiments using EPOCH 2D\citep{EPOCH_Arber2015}
with modified force interpolation scheme by the authors. We use a
pair plasma for simplicity. The time and the spatial coordinates of
the simulations are normalized by the inverse relativistic electron
plasma frequency $\omega_{pe}^{-1}=1/\sqrt{4\pi n_{e}e^{2}/(\gamma_{0}m_{e})}$
and the relativistic electron skin depth $d_{e}=c/\omega_{pe}=c/\sqrt{4\pi n_{e}e^{2}/(\gamma_{0}m_{e})}$,
respectively. We simulate an unmagnetized uniform drifting pair plasma,
which should have no instabilities physically. The instabilities in
the simulations are always numerical artifacts. The simulations for
the unmagnetized uniform drifting pair plasma have been extensively
used in literatures\citep{NCI_Xu2013,NCI_Vay2011} for testing NCI
in PIC codes. We use the method for loading particles in \ref{subsec:load-distribution}.
For the examples we show, we use the WT4 scheme and the regular momentum-conserving
MC4 scheme with $\Delta t\le\Delta x/(2c)$ and $\Delta x_{1}=\Delta x_{2}=\Delta x$.
For Maxwell solvers, we use the one with fourth order dispersion accuracy
(we call it M4), i.e. $\beta_{12}=\beta_{21}=\delta_{1}+1/12=\delta_{2}+1/12=(c\Delta t)^{2}/[12(\Delta x)^{2}]$,
as derived in \ref{sec:dispersion-error} and the Yee solver, i.e.
$\beta_{12}=\beta_{21}=\delta_{1}=\delta_{2}=0$. The parameters we
use for the test problem of a drifting pair plasma are listed in Table
\ref{tab:Parameters-for-test1}.

The growth history of the fraction of the total electromagnetic energy
$E/E_{0}$ is shown in Figure \ref{fig:Results-for-test1}(a) to (d),
where $E_{0}$ is the initial total kinetic energy of all particles
and $E$ is the total energy of electromagnetic field which is a function
of time. The growth of $E/E_{0}$ is always unphysical after an initial
transient that produces electromagnetic fields in thermal equilibrium.
In Figure \ref{fig:Results-for-test1}(a) where we use MC4 scheme
and Yee solver, the case for $\Delta t=\Delta t_{\mathrm{CFL}}/\sqrt{2}$
has slower NCI growth than the cases for $\Delta t\le0.7\Delta t_{\mathrm{CFL}}$,
where $\Delta t_{\mathrm{CFL}}=\Delta x/(\sqrt{2}c)$. This is consistent
with previous studies\citep{NCI_Xu2013,NCI_Vay2011} and can be explained
by the fact that the lowest order NCI growth rate given by Eq(\ref{eq:NCI-Xu2013})
vanishes for $\Delta t=\Delta t_{\mathrm{CFL}}/\sqrt{2}$ but not
for other time steps. In Figure \ref{fig:Results-for-test1}(b) where
we use M4 solver, the case for $\Delta t=\Delta t_{\mathrm{CFL}}/\sqrt{2}$
also has slower NCI growth than other cases, but the time for NCI
to saturate is similar to Yee solver. In Figure \ref{fig:Results-for-test1}(c)
and (d), we use the WT4 scheme. Note that for $\Delta t=\Delta t_{\mathrm{CFL}}/\sqrt{2}$
the WT scheme recovers the standard momentum conserving scheme, and
the scale of time axes for Figure \ref{fig:Results-for-test1}(c)
and (d) is different from that for Figure\ref{fig:Results-for-test1}
(a) and (b). In Figure \ref{fig:Results-for-test1}(c) where we use
WT4 scheme and Yee solver, the NCI grows much slower and saturates
at a much later time if a smaller time step is used. For $\Delta t=0.3\Delta t_{\mathrm{CFL}}$
and $\Delta t=0.1\Delta t_{\mathrm{CFL}}$, $E/E_{0}$ stays between
$10^{-5}$ to $10^{-4}$ for a long time. The results for the tests
with WT4 scheme and M4 solver are shown in Figure \ref{fig:Results-for-test1}(d).
In the case for $\Delta t=0.3\Delta t_{\mathrm{CFL}}$ and $\Delta t=0.1\Delta t_{\mathrm{CFL}}$,
NCI grows slower using M4 solver than using Yee solver. We compare
the color coded plots of the logarithm of the out-of-plane magnetic
field $\log(|\mathrm{FFT}(B_{z})|)$ as a function of wave vector,
for $\Delta t=\Delta t_{\mathrm{CFL}}/\sqrt{2}$ in Figure \ref{fig:Results-for-test1}(e),
and for $\Delta t=0.1\Delta t_{\mathrm{CFL}}$ in Figure \ref{fig:Results-for-test1}(f).
For both plots we use WT4 scheme and M4 solver. The case for $\Delta t=0.1\Delta t_{\mathrm{CFL}}$
has significantly lower numerical instability level than the case
for $\Delta t=\Delta t_{\mathrm{CFL}}/\sqrt{2}$.

The trend observed in the numerical tests using WT4 scheme is that
NCI grows slower if a smaller $\Delta t$ is used, which indicates
that the high order growth rate not included in Eq(\ref{eq:NCI-Xu2013})
depends on $\Delta t$ and decreases as $\Delta t$ decreases. The
detailed analysis of the high order growth rate will be subject of
future reports.

\section{Conclusions\label{sec:Conclusions-and-discussions}}

Using the WT scheme, the lowest order NCI growth rate vanishes if
the spatial derivative stencil in Ampere's equation is not modified
from the standard Yee stencil. The simulation for the drifting pair
plasma becomes more numerically stable when decreasing the time step.
The WT scheme is efficient for improving the quality and flexibility
of relativistic PIC simulations, although the reason for having small
growth rate for small time steps is yet to be understood. The quasi-particle
form-factor used for current deposition is unmodified from the standard
B-spline function, thus good smoothness and negligible fluctuations
at long-range are retained.

\section{Acknowledgement}

Research presented in this paper was supported by the Center for Space
and Earth Science (CSES) program and Laboratory Directed Research
and Development (LDRD) program 20200367ER of Los Alamos National Laboratory
(LANL). The research by PK was also supported by the CSES program.
CSES is funded by LANL's LDRD program under project number 20180475DR.
The simulations were performed with LANL Institutional Computing which
is supported by the U.S. Department of Energy National Nuclear Security
Administration under Contract No. 89233218CNA000001, and with the
Extreme Science and Engineering Discovery Environment (XSEDE), which
is supported by National Science Foundation (NSF) grant number ACI-1548562.
YL and PK are grateful for insightful comments from Dr. Chengkun Huang.

\appendix

\section{Asymptotic NCI growth rate\label{sec:NCI-growth-rate}}

We can calculate the growth rate in Eq(\ref{eq:NCI-Xu2013}) in the
WT scheme by substituting Eq(\ref{eq:shapeEB}) and (\ref{eq:tau_l})
into Eq(\ref{eq:NCI-Xu2013}). We calculate the common factor in $S_{B3}\xi_{0}-S_{E2}[k]_{B1}$
\begin{equation}
\frac{S_{B3}\xi_{0}-S_{E2}[k]_{B1}}{s_{l-1,1}s_{l,2}\tau_{l,3}\eta_{2}}=\cos(\omega^{\prime}\Delta t/2)\bigg(\frac{\sin(k_{1}^{\prime}\Delta x_{1}/2)}{k_{1}^{\prime}\Delta x_{1}/2}\bigg)(-1)^{\nu_{1}}\bigg(\frac{\sin(k_{1}^{\prime}\Delta t/2)}{\Delta t/2}\bigg)-\bigg(\frac{\sin(k_{1}^{\prime}\Delta t)}{k_{1}^{\prime}\Delta t}\bigg)\bigg(\frac{\sin(k_{1}\Delta x_{1}/2)}{\Delta x_{1}/2}\bigg)
\end{equation}
Using the fact that the NCI resonance satisfies the dispersion relation
of the beam $\omega^{\prime}=ck_{1}^{\prime}$, we have
\begin{align*}
\frac{S_{B3}\xi_{0}-S_{E2}[k]_{B1}}{s_{l-1,1}s_{l,2}\tau_{l,3}\eta_{2}} & =\cos(k_{1}^{\prime}c\Delta t/2)\bigg(\frac{\sin(k_{1}^{\prime}\Delta x_{1}/2)}{k_{1}^{\prime}\Delta x_{1}/2}\bigg)(-1)^{\nu_{1}}\bigg(\frac{\sin(k_{1}^{\prime}\Delta t/2)}{\Delta t/2}\bigg)-\bigg(\frac{\sin(k_{1}^{\prime}\Delta t)}{k_{1}^{\prime}\Delta t}\bigg)\bigg(\frac{\sin(k_{1}\Delta x_{1}/2)}{\Delta x_{1}/2}\bigg)\\
 & =\bigg(\frac{\sin(k_{1}^{\prime}\Delta x_{1}/2)}{k_{1}^{\prime}\Delta x_{1}/2}\bigg)(-1)^{\nu_{1}}\bigg(\frac{\sin(k_{1}^{\prime}\Delta t)}{\Delta t}\bigg)-\bigg(\frac{\sin(k_{1}^{\prime}\Delta t)}{k_{1}^{\prime}\Delta t}\bigg)\bigg(\frac{\sin(k_{1}\Delta x_{1}/2)}{\Delta x_{1}/2}\bigg)\\
 & =\frac{\sin(k_{1}^{\prime}\Delta x_{1}/2)(-1)^{\nu_{1}}-\sin(k_{1}\Delta x_{1}/2)}{k_{1}^{\prime}\Delta t\Delta x_{1}/2}
\end{align*}
Using $k_{1}^{\prime}=k_{1}+\nu_{1}(2\pi)/\Delta x_{1}$ we have $\sin(k_{1}^{\prime}\Delta x_{1}/2)\times(-1)^{\nu_{1}}=\sin(k_{1}\Delta x_{1}/2)$,
thus 
\begin{equation}
S_{B3}\xi_{0}-S_{E2}[k]_{B1}=0
\end{equation}
In the same way, we can derive that 
\begin{equation}
S_{B2}\xi_{0}-S_{E3}[k]_{B1}=0
\end{equation}
Thus the NCI growth rate in Eq(\ref{eq:NCI-Xu2013}) is zero for the
WT scheme. The above derivation is valid for arbitrary aliasing beam
and arbitrary spatial derivative stencil in Faraday's equation, as
long as the ultra-relativistic beam is moving along the axis of the
grid, and the spatial derivative stencil in Ampere's equation is not
modified from the standard Yee stencil.

\section{The self-force\label{sec:self-force}}

As long as one uses a charge conserving deposition scheme for calculating
current density, the following Gauss's equation is conserved\citep{Zenitani2015}
\begin{equation}
\hat{\boldsymbol{D}}\cdot\boldsymbol{E}=4\pi\rho\label{eq:gauss}
\end{equation}
where the difference operator $\hat{D}_{1}(X_{n_{1},n_{2},n_{3}})=(X_{n_{1}+\frac{1}{2},n_{2},n_{3}}-X_{n_{1}-\frac{1}{2},n_{2},n_{3}})/\Delta x_{1}$
and similarly for the remaining spatial coordinates. Eq(\ref{eq:gauss})
is conserved automatically, if it is fulfilled in the initial moment.
Following Ref. \citep{1988csup.book.....H}, the approximate equations
used to solve the grid-defined electric fields can be formally expressed
in the form
\begin{equation}
\boldsymbol{E}(\boldsymbol{x}_{g})=V_{c}\sum_{g^{\prime}}\boldsymbol{G}(\boldsymbol{x}_{g};\boldsymbol{x}_{g^{\prime}})\rho(\boldsymbol{x}_{g^{\prime}})
\end{equation}
We assume that the components of the Green's function $\boldsymbol{G}$
satisfies symmetry under the interchange of one coordinate

\begin{equation}
\begin{aligned}G_{1}(\boldsymbol{x}_{g};\boldsymbol{x}_{g^{\prime}}) & =-G_{1}(x_{g^{\prime},1},x_{g,2},x_{g,3};x_{g,1},x_{g^{\prime},2},x_{g^{\prime},3})\\
G_{2}(\boldsymbol{x}_{g};\boldsymbol{x}_{g^{\prime}}) & =-G_{2}(x_{g,1},x_{g^{\prime},2},x_{g,3};x_{g^{\prime},1},x_{g,2},x_{g^{\prime},3})\\
G_{3}(\boldsymbol{x}_{g};\boldsymbol{x}_{g^{\prime}}) & =-G_{3}(x_{g,1},x_{g,2},x_{g^{\prime},3};x_{g^{\prime},1},x_{g^{\prime},2},x_{g,3})
\end{aligned}
\label{eq:green}
\end{equation}
The symmetry can be inherited from the symmetry in the boundary condition,
e.g. periodic boundary condition in each direction. Using the form-factor
in Eq(\ref{eq:shape-real}) and the force interpolation in Eq(\ref{eq:interpolation-EB-real-mod}),
the self-electric-force in $x_{1}$ direction for a particle of charge
$q$ at position $\boldsymbol{x}=(x_{1},x_{2},x_{3})$ gives
\begin{align}
F_{\mathrm{self,1}}(\boldsymbol{x}) & =F_{\mathrm{self,1}}(x_{1},x_{2},x_{3})\nonumber \\
 & =q^{2}\sum_{g,g^{\prime}}G_{1}(\boldsymbol{x}_{g};\boldsymbol{x}_{g^{\prime}})W_{l}(x_{1}-x_{g,1})T_{l}(x_{2}-x_{g,2})T_{l}(x_{3}-x_{g,3})\nonumber \\
 & \qquad\qquad\qquad\qquad\times W_{l}(x_{1}-x_{g^{\prime},1})W_{l}(x_{2}-x_{g^{\prime},2})W_{l}(x_{3}-x_{g^{\prime},3})
\end{align}
Using Eq(\ref{eq:green}) and interchanging $x_{g,1}$ and $x_{g^{\prime},1}$
we have $F_{\mathrm{self,1}}(\boldsymbol{x})=-F_{\mathrm{self,1}}(\boldsymbol{x})$,
thus $F_{\mathrm{self,1}}(\boldsymbol{x})=0$. Similarly $F_{\mathrm{self,1}}(\boldsymbol{x})=F_{\mathrm{self,2}}(\boldsymbol{x})=0$.
The analysis for self-force here only applies to the electrostatic
part of the field. In full electromagnetic PIC, a more comprehensive
analysis for the self-force is desirable.

\section{Dispersion error of the Maxwell solver\label{sec:dispersion-error}}

The dispersion relation of electromagnetic waves for the Maxwell solvers
with modified spatial derivative in Faraday's equation is\citep{Blinne2018}
\begin{equation}
s_{\omega}^{2}=s_{1}^{2}A_{1}+s_{2}^{2}A_{2}+s_{3}^{2}A_{3}
\end{equation}
with the abbreviations
\begin{equation}
s_{\omega}=\frac{\sin(\omega\Delta t/2)}{c\Delta t},\qquad s_{i}=\frac{\sin(k_{i}\Delta x_{i}/2)}{\Delta x_{i}},\quad i=1,2,3
\end{equation}

\begin{equation}
\begin{aligned}A_{1} & =1-2\beta_{12}[1-\cos(k_{2}\Delta x_{2})]-2\beta_{13}[1-\cos(k_{3}\Delta x_{3})]-2\delta_{1}[1-\cos(k_{1}\Delta x_{1})]\\
A_{2} & =1-2\beta_{23}[1-\cos(k_{3}\Delta x_{3})]-2\beta_{21}[1-\cos(k_{1}\Delta x_{1})]-2\delta_{2}[1-\cos(k_{2}\Delta x_{2})]\\
A_{3} & =1-2\beta_{31}[1-\cos(k_{1}\Delta x_{1})]-2\beta_{32}[1-\cos(k_{2}\Delta x_{2})]-2\delta_{3}[1-\cos(k_{3}\Delta x_{3})]
\end{aligned}
\label{eq:Maxwell-A}
\end{equation}
where are $\beta_{ij}$ and $\delta_{i}$ are dimensionless tunable
parameters and the six $\beta$ coefficients depend on $\hat{\beta}_{i}$
as following
\begin{equation}
\hat{\beta}_{1}=\frac{\Delta x_{2}^{2}}{c^{2}\Delta t^{2}}\beta_{12}=\frac{\Delta x_{3}^{2}}{c^{2}\Delta t^{2}}\beta_{13},\qquad\hat{\beta}_{2}=\frac{\Delta x_{3}^{2}}{c^{2}\Delta t^{2}}\beta_{23}=\frac{\Delta x_{1}^{2}}{c^{2}\Delta t^{2}}\beta_{21},\qquad\hat{\beta}_{3}=\frac{\Delta x_{1}^{2}}{c^{2}\Delta t^{2}}\beta_{31}=\frac{\Delta x_{2}^{2}}{c^{2}\Delta t^{2}}\beta_{32}
\end{equation}
In a PIC simulation, if the characteristic wavelength $\lambda$ or
wave vector $k=2\pi/\lambda$ satisfies $k\Delta x=2\pi\Delta x/\lambda\ll1$,
then the dispersion properties near $k\Delta x=0$ is much more important
than the dispersion properties in the rest of the $k$ space. An approach
for minimizing the dispersion error\citep{Blinne2018} can be done
in general cases, but we focus on reducing the dispersion error near
$k\Delta x=0$. We expand the phase velocity $v_{g}=\omega/k$ to
second order using the the spherical coordinates for the wave vectors
$(k_{1},k_{2},k_{3})=(k\sin\theta\cos\phi,k\sin\theta\sin\phi,k\cos\theta)$
\begin{align}
\frac{\omega}{ck} & =1+\bigg[\frac{c^{2}\Delta t^{2}-\Delta x_{1}^{2}(1+12\delta_{1})}{24}\sin^{4}\theta\cos^{4}\phi+\frac{c^{2}\Delta t^{2}-\Delta x_{2}^{2}(1+12\delta_{2})}{24}\sin^{4}\theta\sin^{4}\phi\nonumber \\
 & \qquad+\frac{c^{2}\Delta t^{2}-\Delta x_{3}^{2}(1+12\delta_{3})}{24}\cos^{4}\theta+\frac{c^{2}\Delta t^{2}(1-12\hat{\beta}_{1})}{48}\sin^{2}2\theta\sin^{2}\phi\nonumber \\
 & \qquad+\frac{c^{2}\Delta t^{2}(1-12\hat{\beta}_{2})}{48}\sin^{2}2\theta\cos^{2}\phi+\frac{c^{2}\Delta t^{2}(1-12\hat{\beta}_{3})}{48}\sin^{4}\theta\sin^{2}2\phi\bigg]k^{2}+\mathcal{O}(k\Delta x)^{4}
\end{align}
If we require that the second order term is zero, then we obtain
\begin{equation}
\begin{array}{ccc}
c^{2}\Delta t^{2}-\Delta x_{1}^{2}(1+12\delta_{1})=0 & \qquad & 1-12\hat{\beta}_{1}=0\\
c^{2}\Delta t^{2}-\Delta x_{2}^{2}(1+12\delta_{2})=0 & \qquad & 1-12\hat{\beta}_{2}=0\\
c^{2}\Delta t^{2}-\Delta x_{3}^{2}(1+12\delta_{3})=0 &  & 1-12\hat{\beta}_{3}=0
\end{array}
\end{equation}
which implies
\begin{equation}
\hat{\beta}_{i}=\frac{1}{12},\qquad\delta_{i}=\frac{(c\Delta t/\Delta x_{i})^{2}-1}{12},\qquad\mathrm{and}\ \beta_{ij}=\frac{(c\Delta t/\Delta x_{j})^{2}}{12}
\end{equation}

\section{loading particles with relativistic distributions\label{subsec:load-distribution}}

The method for loading particles with relativistic distributions from
Ref \citep{Zenitani2015} can be generalized to arbitrary drifting
directions. For relativistic PIC simulations, one usually need to
load the pseudo-particles with shifted-Maxwell distribution. The particles
can be loaded in the center-of-mass (CM) frame $S^{\prime}$ where
the distribution function is isotropic and transformed into the simulation
frame $S$, assuming that $S^{\prime}$ is moving at velocity $\overrightarrow{\mathrm{\beta}}c=\overrightarrow{n}\beta c$
with $\beta<1$ w.r.t. $S$, and $\gamma=1/\sqrt{1-\beta^{2}}$. The
commonly used momentum distribution in $S^{\prime}$ frame is usually
the Jüttner-Synga distribution, which represents the thermal equilibrium
state with relativistic temperature $T\apprge mc^{2}/k_{B}$
\begin{equation}
f^{\prime}(\overrightarrow{p}^{\prime})d^{3}\overrightarrow{p}^{\prime}=\frac{N}{4\pi\frac{k_{B}T}{mc^{2}}(mc)^{3}K_{2}\bigg(\frac{mc^{2}}{k_{B}T}\bigg)}\exp\bigg(-\frac{\sqrt{m^{2}c^{2}+p^{\prime2}}}{k_{B}T/c}\bigg)d^{3}\overrightarrow{p}^{\prime}\label{eq:Maxwell-dist-relativistic}
\end{equation}
where $\overrightarrow{p}^{\prime}$ is the momentum of the particle,
$N$ is the number of particles, $m$ is the mass of one particle,
and $K_{2}(x)$ is the modified Bessel function of the second kind.
In low temperature limit $k_{B}T/(mc^{2})\to0$, the distribution
recovers the Maxwell-Boltzmann distribution

\begin{equation}
f^{\prime}(\overrightarrow{p}^{\prime})d^{3}\overrightarrow{p}^{\prime}=\frac{N^{\prime}}{(2\pi mk_{B}T)^{3/2}}\exp\bigg(-\frac{p^{\prime2}}{2mk_{B}T}\bigg)d^{3}\overrightarrow{p}^{\prime}\label{eq:Maxwell-dist-non-relativistic}
\end{equation}
The momentum distribution can be initialized in $S^{\prime}$ using
the widely used Box-Muller algorithm\citep{Box1958} for non-relativistic
Maxwell-Boltzmann distribution in Eq(\ref{eq:Maxwell-dist-non-relativistic}),
or Sobol algorithm\citep{sobol1976} for relativistic Jüttner-Synge
distribution function in Eq(\ref{eq:Maxwell-dist-relativistic}).
The momentum $\overrightarrow{p}^{\prime}$ loaded in $S^{\prime}$
are transformed into momentum $\overrightarrow{p}$ in $S$ frame
by the Lorentz transform
\begin{equation}
\overrightarrow{p}=\bigg[\overrightarrow{p}^{\prime}-(\overrightarrow{p}^{\prime}\cdot\overrightarrow{n})\overrightarrow{n}\bigg]+\gamma\bigg(\overrightarrow{p}^{\prime}\cdot\overrightarrow{n}+\beta\frac{E^{\prime}}{c}\bigg)\overrightarrow{n}
\end{equation}
where $E^{\prime}=\sqrt{p^{\prime2}c^{2}+m^{2}c^{4}}$ is the energy
of the particle in $S^{\prime}$ frame. The momentum distribution
function $f(\overrightarrow{p})$ in $S$ frame is related to $f(\overrightarrow{p}^{\prime})$
by 
\begin{equation}
f(\overrightarrow{p})d^{3}\overrightarrow{p}=\frac{E}{E^{\prime}}f(\overrightarrow{p}^{\prime})d^{3}\overrightarrow{p}^{\prime}=\gamma\bigg(1+\beta c\frac{\overrightarrow{p}^{\prime}\cdot\overrightarrow{n}}{E^{\prime}}\bigg)f(\overrightarrow{p}^{\prime})d^{3}\overrightarrow{p}^{\prime}
\end{equation}
For the volume transform part $\gamma(1+\beta c\overrightarrow{p}^{\prime}\cdot\overrightarrow{n}/E^{\prime})$,
Ref. \citep{Zenitani2015} proposed to use the rejection method. Another
random number $X_{1}\in[0,1]$ is needed to do the rejection. If $(1+\beta c\overrightarrow{p}^{\prime}\cdot\overrightarrow{n}/E^{\prime})/2>X_{1}$,
then we need to reject the pseudo-particle. However, if the particle
distribution in $S^{\prime}$ is symmetric in the $\overrightarrow{n}$
direction, i.e. $f^{\prime}(\overrightarrow{p}^{\prime})=f^{\prime}(\overrightarrow{p}^{\prime}-2\overrightarrow{n}\cdot\overrightarrow{p}^{\prime}\overrightarrow{n})$
and an isotropic distribution in Eq(\ref{eq:Maxwell-dist-relativistic})
or Eq(\ref{eq:Maxwell-dist-non-relativistic}) is a special case for
symmetric distribution, then because $[1+\beta c(\overrightarrow{p}^{\prime}-2\overrightarrow{n}\cdot\overrightarrow{p}^{\prime}\overrightarrow{n})\cdot\overrightarrow{n}/E^{\prime}]/2=(1-\beta c\overrightarrow{p}^{\prime}\cdot\overrightarrow{n}/E^{\prime})/2<1-X_{1}$,
we can flip the momentum $\overrightarrow{p}^{\prime}\to\overrightarrow{p}^{\prime}-2\overrightarrow{n}\cdot\overrightarrow{p}^{\prime}\overrightarrow{n}$
instead of rejecting the pseudo-particle if $(1+\beta c\overrightarrow{p}^{\prime}\cdot\overrightarrow{n}/E^{\prime})/2>X_{1}$.
Then for the symmetric distribution in the $\overrightarrow{n}$ direction,
the acceptance efficiency is $100\%$.

\section{Scaling for ultra-relativistic PIC simulations\label{sec:Scaling-for-relativistic}}

We show that the equations for ultra-relativistic PIC modeling can
be written in the dimensionless form with proper normalization. The
equations for relativistic PIC modeling are

\begin{equation}
\begin{aligned}m_{s}\frac{d\boldsymbol{u}_{s}}{dt} & =q_{s}\bigg(\boldsymbol{E}+\frac{\boldsymbol{v}_{s}}{c}\times\boldsymbol{B}\bigg)\\
\frac{d\boldsymbol{x}_{s}}{dt} & =\boldsymbol{v}_{s}\\
\frac{\partial\boldsymbol{E}}{\partial t} & =c\nabla\times\boldsymbol{B}-4\pi\boldsymbol{J}\\
\frac{\partial\boldsymbol{B}}{\partial t} & =-c\nabla\times\boldsymbol{E}\\
\boldsymbol{J} & =\sum_{s}w_{s}q_{s}\boldsymbol{v}_{s}
\end{aligned}
\end{equation}
where $s$ stands for $s$-th pseudo-particle and $w_{s}$ is the
weight of $s$-th pseudo-particle. We define the normalization
\begin{equation}
\begin{array}{ccccc}
t=\omega_{pe}^{-1}\tilde{t}\qquad & \boldsymbol{x}_{s}=(c/\omega_{pe})\tilde{\boldsymbol{x}}\qquad & \boldsymbol{u}_{s}=\gamma_{0}c\tilde{\boldsymbol{u}}_{s}\qquad & \boldsymbol{v}_{s}=c\tilde{\boldsymbol{v}}_{s}\qquad\\
m_{s}=m_{e}\tilde{m}_{s}\qquad & q_{s}=e\tilde{q}_{s}\qquad & \tilde{w}_{s}=n_{e}\tilde{n}\qquad & \boldsymbol{E}=\sqrt{4\pi\gamma_{0}n_{e}m_{e}c^{2}}\tilde{\boldsymbol{E}}\qquad & \boldsymbol{B}=\sqrt{4\pi\gamma_{0}n_{e}m_{e}c^{2}}\tilde{\boldsymbol{B}}
\end{array}\label{eq:scale-normalization}
\end{equation}
where $\omega_{pe}=\sqrt{4\pi n_{e}e^{2}/(\gamma_{0}m_{e})}$, then
we obtain the dimensionless equations
\begin{equation}
\begin{aligned}\frac{d\tilde{\boldsymbol{u}}_{s}}{d\tilde{t}} & =\frac{\tilde{q}_{s}}{\tilde{m}_{s}}(\tilde{\boldsymbol{E}}+\tilde{\boldsymbol{v}}_{s}\times\tilde{\boldsymbol{B}})\\
\frac{d\tilde{\boldsymbol{x}}_{s}}{d\tilde{t}} & =\frac{\boldsymbol{v}_{s}}{c}\\
\frac{\partial\tilde{\boldsymbol{E}}}{\partial\tilde{t}} & =\tilde{\nabla}\times\tilde{\boldsymbol{B}}-\sum_{s}\tilde{w}_{s}\tilde{q}_{s}\tilde{\boldsymbol{v}}_{s}\\
\frac{\partial\tilde{\boldsymbol{B}}}{\partial\tilde{t}} & =-\tilde{\nabla}\times\tilde{\boldsymbol{E}}
\end{aligned}
\label{eq:scale-eq1}
\end{equation}
and 
\begin{equation}
\tilde{\boldsymbol{v}}=\frac{1}{\sqrt{1+1/\tilde{u}^{2}\gamma_{0}^{2}}}\frac{\tilde{\boldsymbol{u}}}{\tilde{u}}=\bigg[1-\frac{1}{2\tilde{u}^{2}\gamma_{0}^{2}}+\mathcal{O}\bigg(\frac{1}{\tilde{u}^{4}\gamma_{0}^{4}}\bigg)\bigg]\frac{\tilde{\boldsymbol{u}}}{\tilde{u}}\label{eq:sale-eq2}
\end{equation}
If $\gamma_{0}\gg1$ and $\tilde{u}\apprge1$, then we have
\begin{equation}
\tilde{\boldsymbol{v}}=\frac{\tilde{\boldsymbol{u}}}{\tilde{u}}+\mathcal{O}\bigg(\frac{1}{\gamma_{0}^{2}}\bigg)\label{eq:scale-eq3}
\end{equation}
The dimensionless equations Eq(\ref{eq:scale-eq1}) and Eq(\ref{eq:scale-eq3})
are independent of the typical Lorentz factor $\gamma_{0}$ of the
ultra-relativistic system. The scaling relations in Eq(\ref{eq:scale-normalization})
can be used for Lorentz factor scaling of the relativistic PIC simulations,
i.e. the simulation results obtained for one value of $\gamma_{0}$
can be scaled to get the results for other values of $\gamma_{0}$
as long as $\gamma_{0}$ is large.

\section*{References}

\bibliographystyle{elsarticle-num}
\bibliography{pic_method}

\end{document}